\documentclass[universe,article,accept,pdftex]{Definitions/mdpi} 
\usepackage{amssymb}
\usepackage{xcolor}

\usepackage{booktabs} 
\usepackage{multirow}
\usepackage{soul} 
\usepackage{microtype}

\newcommand{\beqn}{\begin{eqnarray}}
\newcommand{\eeqn}{\end{eqnarray}}
\newcommand{\bs}{\boldsymbol}
\newcommand{\eq}[1]{(\ref{#1})}
\newcommand{\avr}[1]{\left\langle #1 \right\rangle}
\newcommand{\cQ}{{\cal Q}}
\newcommand{\cT}{{\cal T}}
\newcommand{\bQ}{{\overline Q}}
\newcommand{\cO}{{\cal O}}

\newcommand{\cZ}{{\cal Z}}
\newcommand{\cP}{{\cal P}}
\newcommand{\dd}{\!{\mathrm d}}
\newcommand{\Tr}{{\mathrm{Tr}}\,}

\newcommand{\corr}[1]{#1}
\newcommand{\add}[1]{#1}
\newcommand{\edit}[1]{#1}
\newcommand{\remove}[1]{}

\firstpage{1} 
\pubvolume{6}
\issuenum{11}
\articlenumber{202}
\pubyear{2020}
\copyrightyear{2020}
\history{Received: 7 October 2020; Accepted: 29 October 2020; Published: 31 October 2020}

\pdfoutput=1

\Title{Conformal Anomaly in Yang-Mills Theory and Thermodynamics of Open Confining Strings} 

\Author{Maxim~N.~Chernodub $^{1,2}$\orcidA{}}

\AuthorNames{Maxim Chernodub}

\address{%
$^{1}$ \quad Institut Denis Poisson UMR 7013, Universit\'e de Tours, 37200 Tours, France;
 maxim.chernodub@idpoisson.fr\\
$^{2}$ \quad Pacific Quantum Center, Far Eastern Federal University, Sukhanova 8, 690950 Vladivostok, Russia}

\abstract{We discuss thermodynamic properties of open confining strings \add{introduced via static sources in the vacuum of} Yang-Mills theory. We derive new sum rules for the chromoelectric and chromomagnetic condensates and use them to show that the presence of the confining string lowers the gluonic pressure in the bulk of the system. The pressure deficit of the gluon plasma is related to the potential energy in the system of heavy quarks and anti-quarks in the plasma.}


\begin{document}

\section{Introduction}

Quantum Chromodynamics (QCD) features many nonperturbative properties thanks to the dynamics of the gluons which mediate the strong fundamental forces. The~pure gluonic sector of the theory is described by Yang-Mills theory which \edit{respects} the conformal invariance at the classical level because the classical Yang-Mills Lagrangian possesses no dimensionful parameters. As~a result, the~classical processes look identically the same under a rescaling of all coordinates, fields, energies, and~momenta according to their canonical~dimensions. 

The conformal symmetry is, however, broken at quantum level. This phenomenon is called quantum conformal anomaly. The~appearance of the conformal anomaly is a natural feature of almost any interacting theory because quantum corrections affect differently the physical phenomena that develop at different energy (or spatial) scales. The~interaction phenomena are determined by coupling constants of the theory that are called ``constants'' in the classical version of the theory. The~presence of the quantum conformal anomaly reveals itself in the fact of the ``running'' of the coupling(s): The~running implies the dependence of the coupling with respect to the energy scale or momentum scale which is involved in the interaction process that is described by this coupling.  Thus, in~an interacting quantum (field) theory, the~coupling constants are not, strictly speaking, constants: they are functions of the energy involved in the interaction~itself.

\edit{The most exciting outcome of experiments on the relativistic heavy-ion collisions is the creation of the quark-gluon plasma~\cite{Gyulassy:2005npa,Muller:2012ann,Jacak:2012nature}.} \add{The thermodynamic and transport phenomena of this deconfined state of strongly interacting matter attract the increasing attention of the scientific community~\cite{Romatschke:2007prl,Heinz:2013ar,Ryu:2015prl}.}

Long time ago it was suggested that an enhanced dissociation of heavy quark-antiquark bound states could be a good signal of the deconfinement phase~\cite{Matsui:1986dk}. Theoretical efforts aimed to study properties of heavy quarkonia require a nonperturbative input in a form of quark-antiquark potentials at finite temperature~\cite{Brambilla:2004jw,Bazavov:2009us}. A~powerful numerical tool to study such potentials is based on first-principle calculations of Polyakov loop correlators in lattice simulations of QCD~\cite{Bazavov:2009us}.
The renormalized correlator of Polyakov loops provides us with the free energy, and---via thermodynamic relations---with internal energy and entropy of the heavy quarks~\cite{Kaczmarek:2002mc}. \add{The Polyakov loops introduce the sources of the chromoelectric field in the vacuum.}
Since \add{the sources of the chromoelectric field are static} they cannot be excited thermally, so that  the
term ``free energy of heavy quarks'' is usually understood as an excess in free energy of thermal gluons 
that appears due to presence of the strong chromoelectric fields of
the heavy quarks~\cite{Kaczmarek:2002mc}.

It is clear that presence of heavy quarks---\add{considered as sources and sinks of the chromoelectric fields}---affects thermodynamics of the quark-gluon plasma. Despite the fact that the thermodynamic effect
of one quark-antiquark pair is not an extensive quantity, a~multiple production of quarkonia may provide a contribution to the
bulk properties of the~plasma.

\add{The presence of the heavy quarks and the emergence of the associated stringy effects were linked in the literature to certain interesting observable effects. The~dilute admixture of heavy quarks was argued in Reference~\cite{Torrieri:2010py} to lower the speed of sound in QGP, while the open string dynamics in a simple model was shown to lead for a thermal-like distribution of transverse mass of particles created in the string decay~\cite{Bialas:1999zg,Steinke:2006cv}.}

\add{In our paper, we show that,} in addition to the free energy, internal energy and entropy of the heavy quarks,
we find that \add{the gluons around the} quarks are also able to make contribution to the pressure of the system. The~effect of pressure is not seen in a
standard approach because the pressure is usually associated with variations of the volume of the system, which does not
enter the excess in the free energy of a finite-sized quark pair. However, below~we show that the variations
in size of a quark system do couple to the~pressure.

In the thermodynamics, the~spatial extent of the quark pair can be treated as a external variable
while the pressure enters a quantity that plays a role of the corresponding generalized force. From~this point of view the renormalized
quark potential can be associated with an excess in a generalized Helmholtz free~energy.

We use a sum rule approach that is generally known as powerful analytical \add{tool} for investigation of certain nonperturbative properties
of QCD physics~\cite{ref:sum:rules:1,ref:sum:rules:2}. \edit{For example, in~absence of the external sources, the~sum rules may help constrain the transport coefficients~\cite{ref:Son}, while the susceptibilities of components of the energy-momentum tensor determine} (derivatives of) thermodynamic potentials~\cite{Meyer:2007fc}.

We derive exact finite-temperature sum rules for excesses in internal energy and in (volume-integrated) pressure in the presence of external
heavy quarks. In~the limit of zero temperature certain combinations of these sum rules reduce to the well-known action and energy sum rules derived by
Michael and Rothe in a lattice formulation of the theory. The~Michael-Rothe equations associate the quark-antiquark potential to an excess in
chromoelectric and chromomagnetic condensates. Historically, the~first attempt to derive such sum rules was done in~Reference~\cite{ref:Michael:1}. The~rules were subsequently corrected~\cite{ref:Rothe:1} and extended~\cite{ref:Michael:2}, and~an important role of the conformal anomaly was stressed~\cite{ref:Rothe:2}.
A lattice check of the sum rules was done in Reference~\cite{ref:Feuerbacher}. Below~we use the new finite-temperature sum rules to study the gluon thermodynamics around color sources~further.

\add{It is important to notice that the ``sum rules'' in the present paper are closer to the finite-temperature generalizations of the ``low-energy QCD theorems''  (similar to the ones developed in References~\cite{Agasian:2002uq,Agasian:2001bj}) as compared to the original sum rules defined in Reference~\cite{ref:sum:rules:1,ref:sum:rules:2}. Here, we follow the lattice terminology, originally used in References~\cite{ref:Michael:1,ref:Rothe:1,ref:Michael:2,ref:Rothe:2}.}

The structure of this paper is as follows. In~Section~\ref{eq:conformal} we give a short introduction to the conformal anomaly in Yang-Mills theory and its relation to the thermodynamics of gluons.  In~Section~\ref{sec:confining:string} we derive the new finite-temperature sum rules that relate the inter-quark-potential to the excesses in expectations values of the chromoelectric and chromomagnetic condensates, that arise due to the presence of the external quark sources. Our conclusions are summarized in the last~section.

\section{The Conformal Anomaly and Thermodynamics of~Gluons}
\label{eq:conformal}
\unskip

\subsection{Conformal Anomaly with Isotropic Renormalization Scale: Equation of State of Gluon~Plasma}

While our paper focuses on the thermodynamics of gluon plasma in the presence of heavy quarks, it is illuminating to discuss first the thermodynamics of Yang-Mills theory without any external quark sources. Although~the derivation is well known, we repeat it for the sake of further~convenience. 

Because of further applications in the lattice gauge theory, we consider a Euclidean formulation of the Yang-Mills theory. The~Euclidean Lagrangian of the theory  with $N$ colors,
\beqn
{\cal L} = \frac{1}{2} \, \Tr G^2_{\mu\nu} \equiv \frac{1}{2} \bigl({\vec {\bs E}}^2 + {\vec {\bs B}}^2\bigr)\,,
\label{eq:cL}
\eeqn
is expressed via the chromoelectric ($E_i^a = G_{i0}^a$ ) and chromomagnetic ($B_i^a = \frac{1}{2} \epsilon_{ijk} G_{jk}^a$) fields, where 
\beqn
G_{\mu\nu} {\equiv} G_{\mu\nu}^a t^a {=} \partial_\mu A_\nu - \partial_\nu A_\mu + i g [A_\mu, A_\nu],
\label{eq:G:mu:nu}
\eeqn
is the field strength tensor of the gluon field $A_\mu \equiv t^a A^a_\mu$. The~generators of the gauge group $t^a$ (with the adjoint index $a=1,\dots,N^2-1$) are normalized in the standard way, $\Tr t^a t^b = \frac{1}{2} \delta^{ab}$. The~bold symbols are used for the vectors in coordinate space while the arrow over a symbol denotes a vector in the color~space.

At the classical level, the~energy momentum tensor of the Yang-Mills theory~Equation \eq{eq:cL}
\beqn
\cT_{\mu\nu} = 2 \Tr \left[G_{\mu\sigma} G_{\nu\sigma} - \frac{1}{4} \delta_{\mu\nu} G_{\sigma\rho} G_{\sigma\rho}\right]\,,
\label{eq:T}
\eeqn
is traceless, $(\cT^\mu_{\ \ \mu})_{\mathrm{cl}} = 0$. This property is an inevitable consequence of the fact that the classical Yang-Mills theory is a \add{conformally-invariant} theory that contains no mass or length scales in its classical~Lagrangian.

\edit{The} conformal invariance is broken at the quantum level. This phenomenon makes the trace of the energy-momentum tensor nonzero:
\beqn
\edit{\avr{\cT^\mu_{\ \ \mu} (x)}} = \avr{\frac{\beta(g)}{2 g} G_{\mu\nu}^a(x) G_{\mu\nu}^a(x)} \equiv \avr{\frac{2 \beta(g)}{g} {\cal L}(x)}\,,
\label{eq:Ttrace}
\eeqn
where 
\beqn
\beta(g) = \frac{\partial \, g (\mu)}{\partial \ln \mu} = - g^3 (b_0 + b_1 g^2 + \dots)\,,
\qquad
b_0 = \frac{11 N}{3 (4\pi)^2}\,, \qquad b_1 = \frac{34 N^2}{3 (4\pi)^4}\,.
\label{eq:beta:function}
\eeqn
is the beta function of Yang-Mills theory expressed via the perturbative coefficients $b_i$. The~beta function determines how the Yang-Mills coupling $g$ changes with the renormalization scale $\mu$ which corresponds to the energy of a given process that involves the gluon interaction via $g = g (\mu)$.  This quantum phenomenon makes the vacuum expectation value of the energy-momentum tensor non-vanishing.

Consider the gluonic system in a finite but large volume $V$ (\add{we} will send it to infinity at the end of calculation). In~the thermodynamic limit, the~total energy of the system is $E = \varepsilon \, V$, where $\varepsilon$ is the energy density. The~total energy $E$ and the pressure $P$ are determined via the derivatives of the partition function $\cZ$ with respect to the temperature $T$ and volume, respectively,
\beqn
E = T \frac{\partial \ln \cZ}{\partial \ln T}\,,
\qquad
P V = T \frac{\partial \ln \cZ}{\partial \ln V}\,,
\qquad
\ln \cZ = - \frac{F}{T} \equiv - f \frac{V}{T}\,,
\label{eq:EP}
\eeqn
and the partition function $\cZ$ is expressed via the free energy $F \equiv f V$.

Since the partition is a dimensionless function of the dimensionful quantities $V$, $T$, and~$\mu$, on~the dimensional grounds one gets the following relation:
\beqn
\Bigl(3 \frac{\partial }{\partial \ln V} - \frac{\partial }{\partial \ln T} - \frac{\partial }{\partial \ln \mu}\Bigr) \ln \cZ = 0\,.
\label{eq:general:lnZ}
\eeqn

The free energy is a physical quantity which characterizes the system as a whole. Therefore, in~the thermodynamic limit, it does not depend on the interaction energy scale~\cite{ref:Dosch}: 
\vspace{6pt}
\beqn
\frac{{\mathrm d} \ln Z}{{\mathrm d} \ln \mu } \equiv \left(\frac{\partial}{\partial \ln \mu}
+ \beta(g) \frac{\partial}{\partial g}\right) \ln Z = 0\,,
\label{eq:renorm:lnZ}
\eeqn
where the beta function is given in Equation~\eq{eq:beta:function}. Then we use Equations~\eq{eq:cL}, \eq{eq:Ttrace}, \eq{eq:EP}, \eq{eq:general:lnZ} and~\eq{eq:renorm:lnZ} to~\edit{identify} 
\beqn
\frac{\partial \ln Z}{\partial g} = \frac{2 V}{T} \left\langle \frac{1}{g} {\cal L} \right\rangle,
\eeqn
and to derive the following conformal anomaly relation:
\beqn
\Theta =\add{\avr{\cT^\mu_{\ \ \mu}}} \equiv E - 3 P V = \int \dd^3 x \, {\Bigl\langle\!\!\Bigl\langle \frac{\beta(g)}{g}
\Bigl[ {\vec {\bs E}}^2(x) + {\vec {\bs B}}^2(x) \Bigr]\Bigr\rangle\!\!\Bigr\rangle}_T. \qquad
\label{eq:anomaly:continuum}
\eeqn

The left-hand side of this equation is related to the equation of state of the gluon plasma while its right-hand side represents the spatial integral of the anomalous trace~Equation \eq{eq:Ttrace} of the energy-momentum tensor~Equation \eq{eq:T}. 

To pick up the thermal effects, Equation~\eq{eq:anomaly:continuum} is regularized via the subtraction of the zero-temperature contribution, with~the notations $\langle\!\langle \dots \rangle\!\rangle_T = \langle \dots \rangle_T - \langle\dots \rangle_{T=0}$. 
\add{Here $\langle \dots \rangle_T$ indicates the expectation value at temperature $T$ while $\langle\dots \rangle_{T=0}$ denotes the expectation value of the same operator at zero temperature.} The reason for this subtraction is that in the perturbation theory, the~vacuum expectation value of the gluon condensate~Equation \eq{eq:Ttrace} diverges quartically with the ultraviolet cutoff. This hazardous divergence is related to the zero-point quantum fluctuations rather than the system's thermodynamics. Since the divergence is a temperature-independent property, it is customary to renormalize the right-hand side of Equation~\eq{eq:anomaly:continuum} by subtracting the divergent $T=0$ part.

The conformal anomaly~Equation \eq{eq:anomaly:continuum} is used in numerical calculations of the equation of state of Yang-Mills theory implemented in first-principle lattice simulations~\cite{Boyd:1996bx, Engels:1988ph}. \edit{Equation~\eq{eq:EP}} give us the total energy and pressure of the gluon plasma, and~determines the quantity \add{on} the right-hand side of~Equation~\eq{eq:anomaly:continuum}:
\beqn
E = F - T \frac{\partial F}{\partial T}\,,
\qquad
P = - \frac{\partial F}{\partial V} \equiv - \frac{F}{V} \equiv - f\,, 
\qquad
\Theta = V T^5 \frac{\partial}{\partial T} \frac{P(T)}{T^4}.
\label{eq:EP:standard}
\eeqn

Thus, the~left hand side of the anomaly Equation~\eq{eq:anomaly:continuum} is expressed via pressure $P$ while the right-hand side \add{of this equation} can be evaluated using numerical calculations in lattice gauge theory. The~full information on the thermodynamics of the theory, $E = E(T)$ and $P = P(T)$ may be obtained via a numerical integration of this simple~equation.
 
In the thermodynamic limit of a uniform system---such as the vacuum of $SU(N)$ gauge theory---the energy density and pressure are extensive variables related to each other via the thermodynamical identities. The~relation of Equation~\eq{eq:anomaly:continuum} provides us with a relationship between the system's energy and pressure to have the complete system of two equations for two unknown quantities. However, the~contributions to energy density and pressure from external sources (from~static quark-antiquark systems) are not extensive variables. Thus, the~standard relation between energy and pressure should not work in these systems by default, and~we need to define these quantities independently. We will perform this task~below.

\subsection{Conformal Anomaly with Anisotropic Renormalization Scales: Energy and Pressure via~Condensates}

Let us generalize the anomaly relation~Equation \eq{eq:anomaly:continuum} by introducing separate renormalization scales in temporal ($\mu = \mu_t$) and spatial ($\mu = \mu_s$) directions. This procedure has been implemented already in the \add{lattice} gauge theory~\cite{Engels:1980ty,Karsch:1982ve}, and~here we will reformulate it in continuum terms for the future use. The~essential point is that the spatial scale $\mu_s$ controls the three-dimensional volume $V$ of the system but does not affect the temperature while the (Euclidean) temporal scale $\mu_t$ controls the temperature $T$ and does not affect the volume. The~usefulness of this separation is evident in view of the thermodynamic relations~Equation \eq{eq:EP:standard}. Then, the~analogues of the dimensional identity~Equation~\eq{eq:general:lnZ} are as follows:
\beqn
3 \frac{\partial \ln \cZ}{\partial \ln V} - \frac{\partial \ln \cZ}{\partial \ln \mu_s} = 0\,,
\qquad
\frac{\partial \ln \cZ}{\partial \ln T} + \frac{\partial \ln \cZ}{\partial \ln \mu_t}= 0\,,
\qquad
\label{eq:separate:lnZ}
\eeqn
while the requirement of the scale independence of the partition function provides us with following counterparts of Equation~\eq{eq:renorm:lnZ}:
\beqn
\frac{{\mathrm d} \ln Z}{{\mathrm d} \ln \mu_{a} } \equiv \frac{\partial \ln Z}{\partial \ln \mu_a}
+ \sum_{b=s,t}\frac{\partial g_b}{\partial \ln \mu_a} \frac{\partial \ln Z}{\partial g_b}  = 0\,, 
\qquad
\label{eq:renorm:lnZ2}
\eeqn
for spatial ($a = s$) and temporal ($a = t$) components. The~spatial and temporal couplings in Equation~\eq{eq:renorm:lnZ2} are defined, respectively, via the relations:
\beqn
\frac{1}{g_s^2(\mu_s,\mu_t)} = \frac{1}{g^2(\mu_s,\mu_t)} \frac{\mu_t}{\mu_s}\,,
\qquad
\frac{1}{g_t^2(\mu_s,\mu_t)} = \frac{1}{g^2(\mu_s,\mu_t)} \frac{\mu_s}{\mu_t}\,.
\label{eq:gs:gt}
\eeqn

The couplings $g_s$ and $g_t$ enter, respectively, the~chromomagnetic ${\vec {\bs B}}$ and chromoelectric ${\vec {\bs E}}$ components of the gluon strength tensor~Equation \eq{eq:G:mu:nu}. 

\add{
While the relations similar to Equation~\eq{eq:gs:gt} appear naturally in the lattice gauge theory on asymmetric lattices~\cite{Karsch:1982ve}, these definitions of the couplings are not obvious in the continuum context which we use in the paper. To~this end, let us briefly adapt our notations to the lattice ones and follow the logic of Reference~\cite{Karsch:1982ve} but in terms of the continuum variables. We rescale the gauge fields \mbox{$A_\mu \to g^{-1} A_\mu$} so that the gluon strength tensor~Equation \eq{eq:G:mu:nu} loses the dependence on the coupling $g$ similarly to the lattice gauge theory. These formulations are equivalent in the continuum limit. The~action of the isotropic theory---with the same scales in the spatial and temporal directions---\mbox{is as follows}:
\beqn
S = \int d^4 x  \left[ \frac{1}{2 g^2} {\vec {\bs B}}^2 + \frac{1}{2 g^2}  {\vec {\bs E}}^2\right] \qquad \mbox{(isotropic)}.
\label{eq:S:iso}
\eeqn

\text{We would like to make the theory invariant} under independent rescaling transformations of spatial, ${\bs x} \to \lambda_s {\bs x}$, and~temporal, $x_4 \to \lambda_t x_4$, coordinates where $\lambda_s$ and $\lambda_t$ are real-valued factors. The~corresponding energy (renormalization) scales transform as $\mu_s \to \lambda_s^{-1} \mu_s$ and $\mu_t \to \lambda_t^{-1} \mu_t$. The~chromoelectric and chromomagnetic fields rescale, respectively, as~follows: ${\vec {\bs E}} \to \lambda^{-1}_s \lambda^{-1}_t {\vec {\bs E}}$ and $ {\vec {\bs B}} \to \lambda^{-2}_s {\vec {\bs B}}$. Since the magnetic and electric parts of the action scale in different manner, the couplings in front of these terms should  not, logically, be the same. In~the theory with anisotropic scales $\mu_s \neq \mu_t$, one gets:
\beqn
S = \int d^4 x  \left[ \frac{1}{2 g^2_s} {\vec {\bs B}}^2 + \frac{1}{2 g^2_t}  {\vec {\bs E}}^2\right] \qquad \mbox{(anisotropic)},
\label{eq:S:ani}
\eeqn
where the couplings $g_s$ and $g_t$ are given in Equation~\eq{eq:gs:gt}, respectively. 

The classical theory~Equation \eq{eq:S:ani} is invariant under the scaling transformations with independent spatial $\lambda_s$ and temporal $\lambda_t$ factors. Since the measure of the integration transforms as $d^4 x \to \lambda_s^3 \lambda_t d^4 x$ we arrive to:
\beqn
d^4 x  \frac{1}{g^2_s} {\vec {\bs B}}^2 & \to & 
(\lambda_t  \lambda_s^3 d^4 x)  \left( \frac{\lambda_s}{\lambda_t} \frac{1}{g^2_s} \right) 
\left( \frac{1}{\lambda_s^4}{\vec {\bs B}}^2 \right) \quad \to d^4 x  \frac{1}{g^2_s} {\vec {\bs B}}^2,
\\
d^4 x  \frac{1}{g^2_t} {\vec {\bs E}}^2 & \to &
(\lambda_t  \lambda_s^3 d^4 x)  \left( \frac{\lambda_t}{\lambda_s} \frac{1}{g^2_t} \right) \left( \frac{1}{\lambda_s^2 \lambda_t^2}{\vec {\bs B}}^2 \right)
\to
d^4 x  \frac{1}{g^2_t} {\vec {\bs E}}^2.
\eeqn

These relations imply that the classical action does not change under the scaling transformation, $S \to S$. The~derivation in lattice terms also uses the naive continuum limit. We refer the interested reader to Reference~\cite{Karsch:1982ve} for the details.
}

In Equation~\eq{eq:gs:gt}, the~quantity $g^2 = g^2(\mu_s,\mu_t)$ is the isotropic Yang-Mills coupling which depends on both scales $\mu_s$ and $\mu_t$. At~the end of our derivations, after~performing all the differentiations, we~always set $\mu_t = \mu_s = \mu$ to achieve a locally isotropic  theory at the single renormalization scale $\mu$.

Then Equations~\eq{eq:separate:lnZ}--\eq{eq:gs:gt} provide us with following expressions for the energy and pressure:
\beqn
\int \dd^3 x \, \varepsilon
& = & \int \dd^3 x \, \Biggl\{{\Bigl\langle\!\!\Bigl\langle \frac{1}{2} \Bigl[ {\vec {\bs B}}^2(x) - {\vec {\bs E}}^2(x) \Bigr]\Bigr\rangle\!\!\Bigr\rangle}_T
+ {\Bigl\langle\!\!\Bigl\langle \frac{\beta_t(g)}{g}
\Bigl[ {\vec {\bs E}}^2(x) + {\vec {\bs B}}^2(x) \Bigr]\Bigr\rangle\!\!\Bigr\rangle}_T \Biggr\}\,,
\label{eq:anomaly:energy}\\
3 \int \dd^3 x \, P
 & = & \int \dd^3 x \, \Biggl\{{\Bigl\langle\!\!\Bigl\langle \frac{1}{2} \Bigl[ {\vec {\bs B}}^2(x) - {\vec {\bs E}}^2(x) \Bigr]\Bigr\rangle\!\!\Bigr\rangle}_T 
 - {\Bigl\langle\!\!\Bigl\langle \frac{\beta_s(g)}{g}
\Bigl[ {\vec {\bs E}}^2(x) + {\vec {\bs B}}^2(x) \Bigr]\Bigr\rangle\!\!\Bigr\rangle}_T\Biggr\}\,,
\label{eq:anomaly:pressure}
\eeqn
where the generalizations of the beta function~Equation \eq{eq:beta:function} are expressed via the isotropic coupling $g^2$:
\beqn
\beta_{a}\bigl(g(\mu)\bigr) = \frac{\partial g (\mu_t,\mu_s)}{\partial \ln \mu_{a}} {\biggl|}_{\mu_s = \mu_t = \mu}\,,
\qquad
a = s, t\,.
\label{eq:beta:functions}
\eeqn

The relations~Equations \eq{eq:anomaly:energy} and \eq{eq:anomaly:pressure} are the continuum analogues of the discretized lattice expressions derived in Reference~\cite{Engels:1980ty}. Notice also that $\langle\!\langle {\vec {\bs B}}^2(x) - {\vec {\bs E}}^2(x) \rangle\!\rangle_T \equiv \langle {\vec {\bs B}}^2(x) - {\vec {\bs E}}^2(x)\rangle_T$ because the zero-temperature contributions disappear in this~term.

The first term in the curvy brackets in each of Equations~\eq{eq:anomaly:energy} and \eq{eq:anomaly:pressure}
is a classical contribution coming from the classical energy-momentum tensor~Equation \eq{eq:T}.
The second term in these expressions corresponds to the quantum correction related to the conformal trace anomaly~Equation \eq{eq:Ttrace}.
Thus, it is natural that the first terms cancel each other in the trace of the energy-momentum tensor,
$\edit{\cT^\mu_{\ \ \mu}} = \varepsilon - 3 P$, while the second terms add up giving us Equation~\eq{eq:anomaly:continuum}. 
At the last step, we used  the equality $\beta_t + \beta_s = \beta$, which \edit{follows} immediately from the definitions~Equation \eq{eq:gs:gt}. The~spatial and temporal beta functions~Equation \eq{eq:beta:functions} were calculated numerically in Reference~\cite{Karsch:1982ve}.

\section{Gluon Energy and Pressure via Gluon Condensates in Presence of Confining~String}
\label{sec:confining:string}
\unskip

\subsection{Thermodynamics from the Conformal~Anomaly}

Now imagine that we have inserted an infinitely heavy quark-antiquark pair in the a thermal vacuum of Yang-Mills theory. Since the quark and the antiquark are so heavy, they cannot be excited by the thermal fluctuations. Nevertheless, the~gluon field around the heavy-quark pair affects the thermal fluctuations of gluons in the thermal bath, and, consequently, \corr{contributes} to energy density and pressure around these static sources of the chromoelectric fields. In~this section, we generalize the thermodynamic relations~Equations \eq{eq:anomaly:energy} and \eq{eq:anomaly:pressure} taking into account the presence of the static~quarks.

How to take into account the influence of the heavy quarks on thermodynamics of gluons?  An~obvious way is to evaluate the gluon condensates in Equations~\eq{eq:anomaly:energy} and \eq{eq:anomaly:pressure} with insertions of quark creation~operators.

A static quark is created at the position $\vec x$ by the gauge-invariant Polyakov loop operator
\beqn
L(\vec x) = \frac{1}{N} \Tr \cP \exp\Bigl\{ i g \int\limits_0^{1/T} \dd t \, A_4(\vec x, t)\Bigr\}\,,
\label{eq:L}
\eeqn
where the operator $\cP$ implies the path ordering of the integration which goes along the straight line parallel to compactified (temperature) direction of the Euclidean space-time. The~length of the compactified direction is $1/T$. The~path of the integration in Equation~\eq{eq:L} is closed via the periodic boundary condition imposed on the gluonic fields in the compactified direction. The~conjugated operator $L^\dagger (\vec x)$ creates an antiquark at the spatial point $\vec x$.

\add{A quantum state corresponding to a system} $\cQ = Q_1 \dots Q_{N_Q} \, \bQ_1 \dots \bQ_{N_\bQ}$, consisting of $N_Q$ quarks located at spatial positions $\vec x_i$ and $N_\bQ$ antiquarks
at points $\vec y_k$, is created by the operator
\beqn
L_\cQ \bigl(\!\{{\vec R}_i\}\bigr) =
L(\vec x_1) \dots  L(\vec x_{N_Q})\cdot L^\dagger(\vec y_1) \dots  L^\dagger(\vec y_{N_\bQ})\,, \quad
\label{eq:L:cQ}
\eeqn
where $\{\vec R_i\} \equiv \{{\vec x_1}, \dots {\vec x_{N_Q}}, {\vec y_1}, \dots {\vec y_{N_\bQ}} \}$. Since
generalizations to various representations is straightforward, we consider here the color-averaged 
operators~Equation \eq{eq:L:cQ} only.

According to Equations~\eq{eq:anomaly:energy} and \eq{eq:anomaly:pressure}, in~order to estimate the effect of the heavy (anti)quark system on the thermodynamics of gluons one should evaluate the chromoelectric and chromomagnetic condensates in the presence of the static (heavy) quarks. Since Equations~\eq{eq:anomaly:energy} and \eq{eq:anomaly:pressure} are linear in the condensates, it is convenient to consider the excess, respectively, in~the energy and in the (integrated-over-volume) pressure caused by the presence of the heavy quarks:
\beqn
E_\cQ(T) = \int \dd^3 x \, \bigl[\varepsilon_\cQ(x) - \varepsilon(x)\bigr]\,,
\label{eq:EcQ}
\qquad
\Pi_\cQ(T) = \int \dd^3 x \, \bigl[P_\cQ(x) - P(x)\bigr]\,.
\label{eq:PcQ}
\eeqn

Notice, that \corr{both these equations} do not depend on the spatial size of the system. These quantities are finite but, in~general, nonzero. 

The contribution of the quark system $\cQ$ \add{to both chromoelectric and chromomagnetic} gluon condensates $\cO$ is given by:
\beqn
\langle \cO \rangle_{\cQ,T} = \frac{\langle \cO \, L_\cQ\rangle_T}{\langle L_\cQ\rangle_T} - \langle \cO \rangle_T\,,
\label{eq:ev:cQ}
\eeqn
where the product $L_\cQ$ of the Polyakov loops is defined in Equation~\eq{eq:L:cQ}. \add{This definition is valid at any nonzero temperature.}

\edit{Using Equations~\eq{eq:anomaly:energy} and \eq{eq:anomaly:pressure} along with \add{the definition in} Equation~\eq{eq:EcQ}}, we get the excesses in \edit{the energy and the pressure} of the system due to the presence of the heavy (anti)quarks
\beqn
 E_\cQ(T)
& = & \int \dd^3 x \, \Biggl\{{\Bigl\langle
\frac{1}{2} \Bigl[ {\vec {\bs B}}^2(x) - {\vec {\bs E}}^2(x) \Bigr]\Bigr\rangle}_{\cQ,T} + {\Bigl\langle \frac{\beta_t(g)}{g}
\Bigl[ {\vec {\bs E}}^2(x) + {\vec {\bs B}}^2(x) \Bigr]\Bigr\rangle}_{\cQ,T} \Biggr\}\,,
\label{eq:anomaly:energy2}\\
3 \,  \Pi_\cQ(T)
& = & \int \dd^3 x \,
\Biggl\{{\Bigl\langle \frac{1}{2} \Bigl[ {\vec {\bs B}}^2(x) - {\vec {\bs E}}^2(x) \Bigr]\Bigr\rangle}_{\cQ,T}
- {\Bigl\langle \frac{\beta_s(g)}{g} \Bigl[ {\vec {\bs E}}^2(x) + {\vec {\bs B}}^2(x) \Bigr]\Bigr\rangle}_{\cQ,T}\Biggr\}\,.
\label{eq:anomaly:pressure2}
\eeqn

Notice that the expectation value $\langle\!\langle \cO \rangle\!\rangle_{\cQ,T}$
in Equations \eq{eq:anomaly:energy} and \eq{eq:anomaly:pressure} gets automatically
replaced by $\langle \cO \rangle_{\cQ,T}$ in Equations \eq{eq:anomaly:energy2} and \eq{eq:anomaly:pressure2} because of cancelations of zero-temperature~contributions.

Equations~\eq{eq:anomaly:energy2} and \eq{eq:anomaly:pressure2} can already be used for a numerical estimation of the contribution of the heavy (anti)quark systems to the thermodynamics of gluons. \add{Yet, there is a way to proceed further~analytically.}

\subsection{Generalization of Michael-Rothe Sum~Rules}

Consider now a quark $Q$ and an antiquark, $\bQ$, separated by the distance $R$. The~color-averaged $Q\bQ$ potential $V_{Q\bQ}$ is given by the following expectation value:
\beqn
e^{- V_{Q\bQ}/T} = \langle L(\vec 0) L^\dagger(\vec R) \rangle\,,
\label{eq:TV}
\eeqn
where we use the symbol $V$ both for the volume (which always appears without a subscript) and for the multi-quark potential $V_\cQ$ (which always comes with a subscript).

Notice that the physical interaction potential of heavy quarks comes from the renormalized Polyakov loops $L^{\mathrm{ren}}(\vec R)$ that are multiplied by a
perimeter/length-dependent factor $Z$ in order to remove the divergent perturbative contribution:
\beqn
L(\vec R) \to L^{\mathrm{ren}}(\vec R) = {[Z(g)]}^{\mu/T} \cdot L(\vec R)\,.
\label{eq:renorm:L}
\eeqn

The renormalization of the quark potential is, however, irrelevant for the determination of both the energy~Equation \eq{eq:anomaly:energy2} and the pressure~Equation \eq{eq:anomaly:pressure2}, because~Equation~\eq{eq:ev:cQ}---used both in Equation~\eq{eq:anomaly:energy2} and in Equation \eq{eq:anomaly:pressure2}---is invariant under
the renormalization shift~Equation \eq{eq:renorm:L}. This property is well understood on physical grounds as the energy~Equation \eq{eq:anomaly:energy2} and pressure~Equation \eq{eq:anomaly:pressure2}  reflect the thermodynamics of gluons around the static quarks and not the quarks~themselves.

The contribution of the gluons around the finite-size heavy-quark system to the thermodynamics is not an extensive quantity. Thus, the~heavy quark potential~Equation \eq{eq:TV} is a function of four~dimensionful parameters: the temperature $T$, the~distance between the quark and antiquark $R$, and~the spatial and temporal renormalization scales $\mu_s$ and $\mu_t$. In~the presence of the heavy quarks, we~rewrite Equation \eq{eq:renorm:lnZ2} and the scale-independence requirements~Equation \eq{eq:separate:lnZ} by using the quantity $\ln \langle L(\vec 0) L^\dagger(\vec R) \rangle \equiv - V_{Q\bQ}/T$ instead of $\ln \cZ$. \edit{Equation~\eq{eq:renorm:lnZ2} reduces to} as follows:
\beqn
\Bigl(\frac{\partial }{\partial \ln R} - \frac{\partial }{\partial \ln \mu_s}\Bigr)
\frac{V_{Q\bQ} (R,T;\mu_s,\mu_t)}{T} = 0\,, \qquad
\Bigl(\frac{\partial }{\partial \ln T} + \frac{\partial }{\partial \ln \mu_t}\Bigr)
\frac{V_{Q\bQ} (R,T;\mu_s,\mu_t)}{T} = 0\,,\qquad
\eeqn
while the second pair of equations in~Equation \eq{eq:renorm:lnZ2} gives us the following identities:
\beqn
T^2 \frac{\partial}{\partial T} \frac{V_{Q\bQ}}{T} & = & - \int \dd^3 x \, \Biggl\{{\Bigl\langle
\frac{1}{2} \Bigl[ {\vec {\bs B}}^2(x) - {\vec {\bs E}}^2(x) \Bigr]\Bigr\rangle}_{{Q\bQ},T} + {\Bigl\langle \frac{\beta_t(g)}{g}
\Bigl[ {\vec {\bs E}}^2(x) + {\vec {\bs B}}^2(x) \Bigr]\Bigr\rangle}_{{Q\bQ},T} \Biggr\}\,,
\label{eq:QQ:energy}\\
R \frac{\partial V_{Q\bQ}}{\partial R}
 & = & - \int \dd^3 x \,
 \Biggl\{{\Bigl\langle \frac{1}{2} \Bigl[ {\vec {\bs B}}^2(x) - {\vec {\bs E}}^2(x) \Bigr]\Bigr\rangle}_{{Q\bQ},T}
- {\Bigl\langle \frac{\beta_s(g)}{g} \Bigl[ {\vec {\bs E}}^2(x) + {\vec {\bs B}}^2(x) \Bigr]\Bigr\rangle}_{{Q\bQ},T}\Biggr\}\,.
\label{eq:QQ:pressure}
\eeqn

Equations \eq{eq:QQ:energy} and \eq{eq:QQ:pressure} are new finite-temperature sum rules that relate
the $Q\bQ$ potential to the excesses in expectations values of the chromoelectric and chromomagnetic condensates,
that arise due~to the presence of the external quark sources. These sum rules represent a generalization of already known low-energy
relations that were derived in $T=0$ lattice gauge theory in References~\cite{ref:Michael:1,ref:Dosch,ref:Rothe:1,ref:Michael:2,ref:Rothe:2}.
For example, subtracting Equation~\eq{eq:QQ:energy} from Equation~\eq{eq:QQ:pressure} and using Equation \eq{eq:L} one gets:
\beqn
\Bigl(1 + R \frac{\partial}{\partial R} - T \frac{\partial}{\partial T} \Bigr) V_{Q\bQ}
= {\Bigl\langle \frac{2 \beta(g)}{g} \int \dd^3 x\, {\cal L}(x)\Bigr\rangle}_{Q\bQ,T}\qquad
\label{eq:fff}
\eeqn

This relation is nothing but a natural finite-temperature extension of a well-known ``action sum rule'' first found in the lattice formulation in Reference~\cite{ref:Rothe:1}:
\beqn
V_{Q\bQ} + R \frac{\partial V_{Q\bQ}}{\partial R}
= {\Bigl\langle \frac{2 \beta(g)}{g} \int \dd^3 x\, {\cal L}(x)\Bigr\rangle}_{Q\bQ (T=0)}\,.\qquad
\label{eq:MR:action}
\eeqn

Moreover, the~zero-temperature limit of Equation~\eq{eq:QQ:energy} gives us back a continuum version of the known ``energy sum rule''~\cite{ref:Rothe:1}:
\beqn
V_{Q\bQ} & = & \int \dd^3 x \, \Biggl\{{\Bigl\langle
\frac{1}{2} \Bigl[ {\vec {\bs B}}^2(x) - {\vec {\bs E}}^2(x) \Bigr]\Bigr\rangle}_{Q\bQ(T=0)}
+ {\Bigl\langle \frac{\beta_t(g)}{g}
\Bigl[ {\vec {\bs E}}^2(x) + {\vec {\bs B}}^2(x) \Bigr]\Bigr\rangle}_{Q\bQ(T=0)} \Biggr\}\,.
\label{eq:QQ:energy:0}
\eeqn

Let us mention that we have three pairs of equations that look similar but have rather different physical~meanings:
\begin{itemize}[leftmargin=21pt,labelsep=7pt]
\item[(i)] The pair Equations \eq{eq:anomaly:energy} and \eq{eq:anomaly:pressure} is an \textit{equality}, that relates the energy and pressure of the gluonic fields to the expectation values of the gluon condensates. This pair is a continuum version of the corresponding lattice formulae derived in Reference~\cite{Engels:1980ty}.

\item[(ii)] The pair Equations \eq{eq:anomaly:energy2} and \eq{eq:anomaly:pressure2} is a natural \textit{definition}
of the contribution of heavy quarks to the energy~Equation \eq{eq:anomaly:energy} and pressure~Equation \eq{eq:anomaly:pressure}
of the~gluons.

\item[(iii)] Finally, the~pair Equations \eq{eq:QQ:energy} and \eq{eq:QQ:pressure} describes the new finite-temperature \textit{sum rules}.
\end{itemize}

\subsection{New Sum Rules and Exact Relations for Gluon Thermodynamics around Static~Sources}
A comparison of the exact relations~Equation \eq{eq:anomaly:energy2}, Equation \eq{eq:anomaly:pressure2}
with new sum rules Equation \eq{eq:QQ:energy}, Equation \eq{eq:QQ:pressure} shows that
the excess in the energy~Equation \eq{eq:EcQ} and the (volume-integrated) excess in the pressure~Equation \eq{eq:PcQ} due to the presence
of the heavy quark-antiquark pair are, respectively:
\beqn
 E_{Q\bQ}(R,T) = V_{Q\bQ} - T \frac{\partial V_{Q\bQ}}{\partial T} \equiv - T^2 \frac{\partial}{\partial T} \frac{V_{Q\bQ}}{T}\,,
\qquad
 \Pi_{Q\bQ}(R,T) & = & - \frac{R}{3} \frac{\partial V_{Q\bQ}}{\partial R}\,.
\label{eq:EP:exact}
\eeqn

The important property of these relations is that they are~exact.

From Equation~\eq{eq:EP:exact} we get a Maxwell-type relation between the excesses in the energy and~pressure:
\beqn
R {\left(\frac{\partial \,  E_{Q\bQ}}{\partial R}\right)}_T
= 3 T^2 {\left(\frac{\partial }{\partial T} \frac{ \Pi_{Q\bQ}}{T}\right)}_R\,,
\label{eq:EoS}
\eeqn
where the subscripts mean that the derivatives are taken at the fixed temperature $T$ and
the fixed inter-quark distance $R$, respectively. Equation~\eq{eq:EoS} is an exact equation of state for heavy~quarks.

A comparison of the energy excess in Equation~\eq{eq:EP:exact} with the corresponding thermodynamical relation [the first formula in~Equation \eq{eq:EP:standard}] would lead us to the conclusion
that the heavy quark potential $V_{Q\bQ}$ may be interpreted---up to a renormalization constant~Equation \eq{eq:renorm:L}---as an excess in the Helmholtz free energy $ F_{Q\bQ}$ due to the presence of the heavy quarks. However, the~excess in the
pressure~Equation \eq{eq:EP:exact} has nothing to do with its analogue in the thermodynamical limit
[the~second formula in~Equation \eq{eq:EP:standard}], making the free energy interpretation of the heavy quark potential $V_{Q\bQ}$~obscure.

Nevertheless, one can still interpret $V_{Q\bQ}$ as an excess in the Helmholtz free energy of gluons due to the presence
of the heavy (anti)quarks. Indeed, besides~the temperature~$T$, the~heavy quark thermodynamics has the additional external
variable $R$ but lacks the usual volume variable $V$. In~this case the usual fundamental
thermodynamic relation, $\dd E = T \, \dd S - P \, \dd V$, should be written as follows
\beqn
\dd E_{Q\bQ} = T \, \dd S_{Q\bQ} - X_{Q\bQ} \, \dd R\,,
\label{eq:fund}
\eeqn
where $S_{Q\bQ}$ is the excess in the entropy and $X_{Q\bQ}$ is the generalized force associated with the distance $R$
between the quark and antiquark. From~Equation~\eq{eq:EP:exact} one gets:
\beqn
 S_{Q\bQ} = - {\left(\frac{\partial V_{Q\bQ}}{\partial T} \right)}_R\,,
\qquad
 X_{Q\bQ} = - {\left(\frac{\partial V_{Q\bQ}}{\partial R} \right)}_T \equiv - \frac{3}{R}  \Pi_{Q\bQ}\,,
\quad
\label{eq:SXQQ}
\eeqn
so that the generalized force $X_{Q\bQ}$ is related to the (integrated) pressure in Equation~\eq{eq:EP:exact}.
Then the differential of the excess in the Helmholtz free energy $F_{Q\bQ}\equiv E_{Q\bQ} - T S_{Q\bQ}$ is:
\beqn
\dd F_{Q\bQ} = -  S_{Q\bQ} \ \dd T -  X_{Q\bQ} \ \dd R\,. \qquad
\label{eq:dF}
\eeqn

Notice that the excess in any thermodynamical quantity coming as a result of the presence of a finite number of finitely-separated quarks is not an extensive property of the system. Due to the effect of screening (i.e., due to a finite mass gap in the spectrum), the~finite-volume corrections to these quantities should  \add{vanish with the increase in the volume of the system}. Thus, the~volume is a thermodynamically irrelevant variable for large enough~systems.

The excess in the free energy $F_{Q\bQ}$ can be easily obtained by an integration of Equation~\eq{eq:dF}:
\beqn
F_{Q\bQ}(R,T) = V_{Q\bQ}(R,T)\,.
\label{eq:F:new:old}
\eeqn

A free integration constant---which inevitably appears in any integration---is to be interpreted
as the renormalization quark-antiquark potential~Equation \eq{eq:renorm:L}. We have just shown that Equation~\eq{eq:F:new:old}---despite its standard appearance---has somewhat nonstandard interpretation
because it involves the generalized fundamental thermodynamic relation~Equation \eq{eq:fund} that includes the
generalized force~Equation \eq{eq:SXQQ}.

\subsection{Negative Pressure~Excess}

The excess in the pressure due to the presence of the heavy quark pairs is always negative.
 This~property is seen from the second formula in Equation~\eq{eq:EP:exact} taking into account that the $V_{Q\bQ}$ potential is a concave function of the distance~\cite{Bachas:1985xs}.
Thus, the~presence of a heavy quarkonium state should decrease the pressure in the gluon plasma. A~gas of weakly-interacting heavy-quark bound states with the density $\rho$ should produce the total pressure deficit in the gluonic plasma (for the sake of simplicity, here we assume that all $Q\bQ$-states have the same size~$R$):
\beqn
\Delta P_{Q\bQ}(R,T) = - \frac{\rho(R,T) R}{3} \frac{\partial V_{Q\bQ}(R,T)}{\partial R} < 0\,.
\eeqn

It is interesting to notice that even at zero temperature the integrated pressure in Equation~\eq{eq:EP:exact} is not a \add{positive-definite} quantity. This fact does not contradict the common wisdom. For~example, a~similar feature characterizes the conventional Casimir effect \add{in the vacuum of electrodynamics}: the~vacuum between two perfectly conducting metallic plates has a negative pressure, negative free energy and negative internal energy~\cite{ref:Casimir:1,ref:Casimir:2}. 
\add{Less exotic example of a similar system at nonzero temperature is a solution of ethanol in water, that demonstrates in normal conditions an effect of volume contraction (i.e., negative excess of volume). The~maximal volume contraction in this binary mixture is reached at approximately 20\% of mole fraction of ethanol~\cite{ref:ethanol}, thus mimicking, in~very loose terms, a~noticeable (excess of) negative pressure.}

Our calculations are in line \edit{with} the observation that the contribution of the heavy quarks to the specific heat of the gluon plasma becomes negative at high enough temperatures~\cite{Noronha:2010hb}. The~identification~Equation \eq{eq:F:new:old} \edit{represents} the excess in the free due to heavy quarks which is equal to the (renormalized) heavy-quark potential (see, e.g.,~Reference~\cite{Kaczmarek:2002mc}). Moreover, there is no {ab~initio} restriction on the sign of the excess in any thermodynamic quantity due to presence of external~particles.

\subsection{Multiquarks and a Single~Quark}
A generalization of our results to multiquark systems is \edit{rather straightforward}.
Instead of the $Q\bQ$ potential $V_{Q\bQ}$ one should use its multiquark counterpart $V_{\cQ}$
defined via the multiquark loop~Equation~\eq{eq:L:cQ} as follows:
\beqn
e^{- V_{\cQ}(\!\{{\vec R}_i\},T)/T} = \langle L_\cQ \bigl(\!\{{\vec R}_i\}\bigr) \rangle\,.
\label{eq:multi:1}
\eeqn

In addition, the~derivatives with respect to the $Q\bQ$ distance should be replaced by the following scale derivative:
\beqn
R \frac{\partial}{\partial R} V_{Q\bQ}(R,T) \to
\frac{\partial}{\partial \lambda} V_\cQ(\{\lambda \vec R_i\},T) {\biggl{|}}_{\lambda = 1}\,.
\label{eq:multi:2}
\eeqn

According to Equations~\eq{eq:EP:exact}, \eq{eq:multi:1} and~\eq{eq:multi:2}, a~single quark~$Q$ cannot affect the pressure of the system even in the deconfinement phase \add{because the distance variable $R$ is evidently absent. Therefore, the~contribution to the vacuum gluonic pressure from the infinite string open at one end is zero.} Then~\add{the single-quark contribution depends only on temperature, and} we get the following relation, which is a direct analogue of Equation~\eq{eq:QQ:pressure} for $\cQ = Q$:
\beqn
& & {\Bigl\langle \Bigr(\frac{1}{2} - \frac{\beta_s(g)}{g} \Bigl) \int \dd^3 x \, {\vec {\bs B}}^2(x)\Bigr\rangle}_{Q,T} =
 {\Bigl\langle \Bigr(\frac{1}{2} + \frac{\beta_s(g)}{g} \Bigl) \int \dd^3 x \, {\vec {\bs E}}^2(x) \Bigr\rangle}_{Q,T}\,.
\label{eq:Q:pressure}
\eeqn

This relation leads to an analogue of Equation~\eq{eq:fff} for a single quark:
\beqn
V_{Q} - T \frac{\partial V_{Q}}{\partial T}
= {\Bigl\langle \frac{2 \beta(g)}{g} \int \dd^3 x\, {\cal L}(x)\Bigr\rangle}_{Q,T}\,.\qquad
\label{eq:Q:energy}
\eeqn

The above equation represents yet another new sum rule, that reduces to the known relation~Equation \eq{eq:MR:action} in the limit $T \to 0$ (to verify the correspondence,  one should notice \mbox{that $\partial V_Q/\partial R \equiv 0$)}.

\subsection{A Few Phenomenological~Examples}

The excesses in free energy~Equation \eq{eq:F:new:old}, entropy~Equation \eq{eq:SXQQ} and internal energy~Equation \eq{eq:EP:exact} were thoroughly studied in lattice simulations of $SU(3)$ Yang-Mills theory~\cite{Kaczmarek:2002mc,Kaczmarek:2005ui}. Here we discuss briefly the new quantity, the~(volume-integrated) excess in the pressure~Equation \eq{eq:EP:exact} caused by the presence of the heavy quark-antiquark~pair.

\subsubsection{Deconfinement~Phase}
The deconfinement phase, \edit{which} is realized at temperatures higher than the critical temperature $T_c$, the~\add{real part of the} heavy-quark potential can \edit{quantitatively} be described by the following formula~\cite{ref:Heavy:Quark}:
\beqn
V_{Q\bQ}(R,T) = - \frac{e(T) T}{(R T)^{d(T)}} \, e^{- \mu(T) R}\,, \qquad [T \geqslant T_c]\,, \quad
\label{eq:V:finite:T}
\eeqn
where $e(T)$ and $d(T)$ are dimensionless temperature-dependent parameters, and~$\mu(T)$ is the screening mass. For~a reference, the~behaviour of the screening factor is $\mu(T) \simeq 2.5 T$ at $T \gtrsim 1.5 T_c$; it decreases drastically as temperature approaches its critical value $T_c$ from above. The~exponent $d$ increases from $d \simeq 1$ at $T = T_c$ to a stable value $d \simeq 1.5$ at $T \gtrsim 1.5 T_c$ up to temperatures $T \approx 3 T_c$~\add{\cite{ref:Heavy:Quark}}.
\add{The heavy-quark potential develops also an imaginary part~\cite{Laine:2006ns}, which may become important for the dynamics of the bottomonium bound state at temperatures around 250 MeV and higher. While the imaginary part of the potential does not enter our analysis below, we mention that its presence reflects the fact that heavy quarkonia are unstable states with finite widths.} \add{We refer the Reader to} References~\cite{Wolschin:2020qxa,Bazavov:2020teh} for~recent reviews \add{on the subject}.

Relation~Equation \eq{eq:EP:exact} implies that in the deconfinement phase the absolute value of the $Q\bQ$-induced pressure is a decreasing function of the distance~$R$:
\beqn
\Pi_{Q\bQ}(R,T\geqslant T_c) = \frac{d + \mu(T) R}{3} V_{Q\bQ}(R,T)<0\,.\quad
\label{eq:P:1}
\eeqn

At large separations between the sources, the~pressure excess vanishes due to the color~screening.

\subsubsection{Confinement Phase at Finite~Temperature}

In the confinement phase at finite-temperature $0 < T < T_c$, the~interaction potential of quarks in the quarkonium state can be described, at~large distances, by~the following formula~\cite{ref:Heavy:Quark}:
\beqn
V_{Q\bQ} = V_0 + \sigma(T) R + C T \ln(2 R T)\,,\qquad R T \gg 1, \qquad [T < T_c]\,,
\eeqn
where $V_0$ and $C$ are parameters obtained by a fit of the lattice numerical data. We get from Equation~\eq{eq:EP:exact}:
\beqn
\Pi_{Q\bQ}(R,T < T_c) = - \frac{C T + \sigma R}{3}\,.
\label{eq:P:2}
\eeqn

At large distances, the~pressure deficit is a linearly decreasing function of the separation between the quark and the~antiquark.

\subsubsection{Zero~Temperature}

At zero temperature, the~heavy-quark interactions at short and long distances can be described by the single Cornell potential:
\beqn
V_{Q\bQ}(R,T=0) = - \frac{\alpha}{R} + \sigma R\,,
\eeqn
where $\alpha \simeq \edit{0.356 \pm 0.015}$ and $\sigma \simeq \edit{(0.207 \pm 0.088)}\, {\mathrm{GeV}}^2$ are the phenomenological constants which may be obtained, \text{for example}, from~the heavy quarkonia \text{(bottomonia)} fits~\cite{eq:Cornell,Mateu:2018zym}). The~integrated pressure~deficit,
\beqn
\Pi_{Q\bQ}(R,T = 0) = - \frac{\alpha}{3 R} - \frac{\sigma R}{3}\,,
\label{eq:P:3}
\eeqn
takes its maximum $\Pi_{Q\bQ}^{\mathrm{max}} = - 2 \sqrt{\alpha \sigma}/3 \simeq - \edit{(178 \pm 34)}\, {\mathrm{MeV}}$ at the distance \add{between the heavy quarks} $R^{\mathrm{max}} = \sqrt{\alpha / \sigma} \simeq \edit{(0.28 \pm 0.06)}\, {\mathrm{fm}}$.

In the confinement phase the integrated excess in the gluon pressure, Equations~\eq{eq:P:2} and \eq{eq:P:3}, is~always a negative quantity, as~expected. At~small distances $R$, the~effect appears due to perturbative gluons in the chromoelectric flux which spans between the~sources.

The pressure deficit~Equation \eq{eq:P:3} decreases linearly with the increase in the $Q\bQ$ separation $R$ at large enough $R$. Physically, this effect emerges due to \edit{the presence}  of the QCD string that is spanned between the quark and antiquark. \edit{The presence} of the string causes the backreaction of the gluons in a close~proximity. 

\section{Conclusions}

We derived the finite-temperature sum rules Equations \eq{eq:QQ:energy} and \eq{eq:QQ:pressure} of gluons in heavy quarks systems. From~our sum rules, the~Michael-Rothe sum rules for the action~Equation \eq{eq:MR:action} and for the energy~Equation \eq{eq:QQ:energy:0} are recovered automatically in the zero-temperature limit. In~addition, the~new sum rules provide us with \text{the new expression}~Equation \eq{eq:EP:exact} for the internal energy $E$ of the quarks~Equation \eq{eq:EcQ}.

At the same time, these sum rules lead to the new expression~Equation \eq{eq:EP:exact} for the spatial integral~Equation \eq{eq:EcQ} of the excess in the gluonic pressure, $\Pi_\cQ$ around quarks. The~excess corresponds to the difference between the pressure in the gluonic system in the presence of the quarks and without quarks. In~the confining phase, this excess may be interpreted as the effect of the confining string, which is open at the (anti-)quarks that serve as sources and sinks of the chromoelectric field. We~found that the pressure excess of the gluons around a heavy quark-antiquark state is always negative, independent of the phase of the gluonic~system.

We also derived the exact equation of state~Equation \eq{eq:EoS} that relates the excess in the gluonic energy to the excess in the  gluonic pressure around the heavy-quark systems. The~generalization of our results to multi-quark systems is straightforward and given in Equations~\eq{eq:multi:1} and \eq{eq:multi:2}.

\funding{The work was partially supported by Grant No. 0657-2020-0015 of the Ministry of Science and Higher Education of~Russia.}

\acknowledgments{The author is grateful to Stam Nicolis for useful~discussions.}
\conflictsofinterest{The author declares no conflict of interest.} 
\reftitle{References}

\vspace{6pt}\noindent{\fontsize{9}{9}\selectfont\textbf{Publisher's Note :} { MDPI stays neutral with regard to jurisdictional claims in published maps and institutional affiliations.}\par}
\end{document}